\def\year{2020}\relax
\documentclass[letterpaper]{article} 
\usepackage{aaai20}  
\usepackage{times}  
\usepackage{helvet} 
\usepackage{courier}  
\usepackage{url}  
\usepackage{graphicx} 
\usepackage{graphicx}  
\usepackage{amsmath}
\usepackage{amssymb}
\usepackage{xcolor}
\usepackage{multirow}
\usepackage{subfigure}
\title{RIS-GAN: Explore Residual and Illumination with Generative Adversarial Networks for Shadow Removal}
\author{Ling Zhang\textsuperscript{\rm 1},
Chengjiang Long\textsuperscript{\rm 2}\thanks{This work was co-supervised by Chengjiang Long and Chunxia Xiao.},
Xiaolong Zhang\textsuperscript{\rm 1}
Chunxia Xiao\textsuperscript{\rm 3}$^*$\\
\textsuperscript{\rm 1}Wuhan University of Science and Technology, Wuhan, Hubei, China\\
\textsuperscript{\rm 2}{Kitware Inc. Clifton Park, NY, USA} \\
\textsuperscript{\rm 3}{School of Computer Science, Wuhan University, Wuhan, Hubei, China} \\
\{zhling, xiaolong.zhang\}@wust.edu.cn, chengjiang.long@kitware.com, cxxiao@whu.edu.cn
}

\begin{document}

\maketitle

\begin{abstract}
Residual images and illumination estimation have been proved very helpful in image enhancement. In this paper, we propose a general and novel framework RIS-GAN which explores residual and illumination with Generative Adversarial Networks for shadow removal. Combined with the coarse shadow-removal image, the estimated negative residual images and inverse illumination maps can be used to generate indirect shadow-removal images to refine the coarse shadow-removal result to the fine shadow-free image in a coarse-to-fine fashion. Three discriminators are designed to distinguish whether the predicted negative residual images, shadow-removal images, and the inverse illumination maps are real or fake jointly compared with the corresponding ground-truth information. To our best knowledge, we are the first one to explore residual and illumination for shadow removal. We evaluate our proposed method on two benchmark datasets, {\em i.e.}, SRD and ISTD,  and the extensive experiments demonstrate that our proposed method achieves the superior performance to state-of-the-arts, although we have no particular shadow-aware components designed in our generators. Our source code is available at {\em \color{red}{\url{https://github.com/zhling2020/RIS-GAN}}}.
\end{abstract}

\noindent

\section{Introduction}
Shadow is a ubiquitous natural phenomenon, which is appeared when the light is partial or complete blocked, bringing down the accuracy and effectiveness of some computer vision tasks, such as target tracking, object detection and recognition\cite{Miki2000Moving,long2014accurate,Cucchiara2002Improving,hua2013collaborative,long2015multi,Long2017CVPR,hua2018collaborative,luo2019end}, image segmentation and intrinsic image decomposition
\cite{Li2018Learning}. Therefore, it is necessary to conduct shadow removal to improve the visual effect of image and video editing, such as film and television post-editing. However, it is still a very challenging problem to 
remove shadow in complex scenes due to illumination change, texture variation, and other environmental factors. 

A variety of existing works including traditional methods\cite{Shor2008The,Xiao2013Fast,zhang2019effective}, and learning-based methods \cite{Gryka2015Learning,Wang2017Stacked,le2018a} have been developed to solve this challenging problem. Different from traditional methods that highly rely on some prior knowledge ({\em e.g.}, constant illumination and gradients) and often bring obvious artifacts on the shadow boundaries, learning-based methods especially recent deep learning methods like \cite{Hu2018Direction} and \cite{Sidorov2018Conditional} have achieved some advances.
However, the effectiveness of these methods highly depends on the training dataset and the designed network architectures. When the training set is insufficient or the network model is deficient, they are insufficient to produce desired shadow detection masks and shadow-removal images. 
Also, most of existing deep learning methods just focus on shadow itself, without well exploring other extra information like residual and illumination for shadow removal.

\begin{figure*}[ht!]
  \centering
  \includegraphics[width=0.85\linewidth]{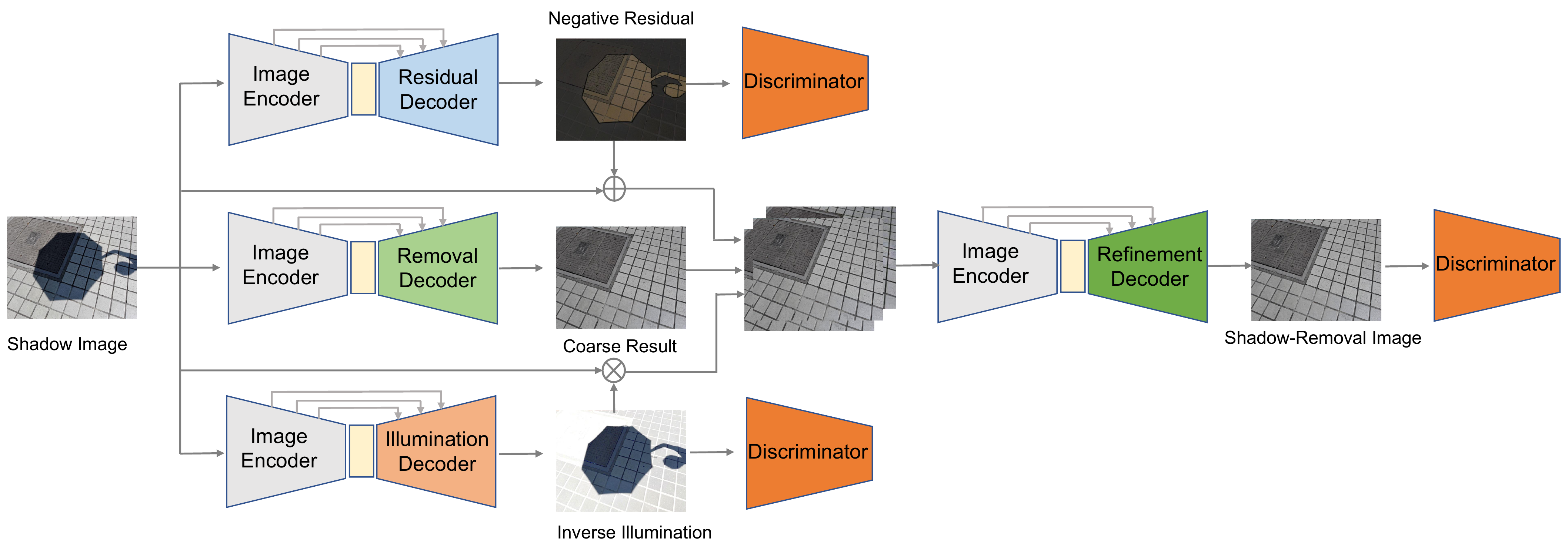}
  \vspace{-0.3cm}
  \caption{The framework of our proposed RIS-GAN, which is composed of four generators in the encoder-decoder structure and three discriminators. It takes a shadow image as input and outputs the negative residual image, inverse illumination map, and shadow-removal image in an end-to-end manner. Note that these four generators share the same architecture, and the three discriminators share the same parameters. No particular shadow-aware components are designed in the whole framework. }
\label{fig:pipeline}
\end{figure*}

In this paper, we propose a general framework RIS-GAN to explore both residual and illumination with Generative Adversarial Networks for shadow removal, unlike Sidorov's AngularGAN \cite{Sidorov2018Conditional} which just introduces an illumniation-based angular loss without estimating illumination color or illumination color map. As illustrated in Figure \ref{fig:pipeline}, our RIS-GAN consists of four generators in the encoder-decoder structure and three discriminators. Such four generators are designed to generate negative residual images, intermediate shadow-removal images, inverse illumination maps, 
and refined shadow-removal images. In principle, unlike the existing deep learning methods \cite{Qu2017DeshadowNet,Hu2018Direction,zhu2018bidirectional} which are designed particular for shadow-removal, any kinds of encoder-decoder structures can be used as our generators.

For the residual generator, we follow the idea of negative residual \cite{fu2017removing} and let the generator take a shadow image to generate a negative residual for detecting shadow area and recover a shadow-lighter or shadow-free image by applying a element-wise addition with the input shadow image indirectly. For the illumination generator, we design it based on the Retinex model \cite{fu2016weighted,guo2016lime,wang2019underexposed} where the ground-truth shadow-free image can be considered as a reflectance image, and the shadow image is the observed image.
The output of the illumination generator is a inverse illumination map which can be used for shadow region detection and recovering a shadow-removal image by applying a element-wise multiplication with the input shadow image indirectly.

We shall emphasize that we use the refinement generator to refine the shadow-removal images obtained from the shadow removal generator in a coarse-to-fine fashion. Besides the coarse shadow-removal results, we also incorporate two indirect shadow-removal images via the explored negative residual and inverse illumination to recover the final fine shadow-removal image. With this treatment, the shadow-removal refinement generator has three complementary input sources, which ensures the high-quality shadow removal results.

Like all the Generative Adversarial Networks (GANs), our proposed RIS-GAN adopts the adversarial training process \cite{goodfellow2014generative} between the four generators and three discriminators alternatively to generate high-quality negative residual images, inverse illumination maps, 
and a shadow-removal images. It is worth mentioning that we design a cross loss function to make sure the recovered shadow-removal image is consistent with the explored residual and illumination. We also adopt a joint discriminator \cite{He2018Densely} to ensure that three discriminators share the same architecture with the same parameter values to judge whether the generated results are real or fake compared with the corresponding ground-truth, which can make sure all the produced results are indistinguishable from the corresponding ground-truth. With the number of epochs increases, both generators and discriminators improve their functionalities so that it becomes harder and harder to distinguish a generated output from the corresponding ground-truths. Therefore, after a certain large number of training epochs, we can utilize the learned parameters in the generators to generate a negative residual image, a inverse illumination map, 
and a shadow-removal image.

Different from the existing shadow detection and removal methods, our main contributions can be summarized as three-fold: 
(1) we are the first one to propose a general and novel framework RIS-GAN 
    with generators in an encoder-decoder structure to explore residual and illumination between shadow and shadow-free images for shadow removal;
(2) the correlation among residual, illumination and shadow has been well explored within the cross loss function and the joint discriminator 
    and we are able to get complementary input sources for better improving the quality of shadow-removal results;  
and (3) without any particular shadow-aware components in our encoder-decoder structured generators, our proposed RIS-GAN still achieves the outperformance to art-of-the-arts. Such experimental resuts clearly demonstrates the efficacy of the proposed approach.

\section{Related Work}
Shadow removal is to recover a shadow-free image. One typical group of traditional methods is to recover the illumination in shadow regions using illumination transfer \cite{Xiao2013Efficient,Guo2011Single,Zhang2015Shadow,Khan2016Automatic}, which borrow the illumination from non-shadow regions to shadow regions. 
Another typical ground of traditional methods involves gradient domain manipulation \cite{Finlayson2005On,Liu2008Texture,Feng2008Texture}. 
Due to the influence of the illumination change at the shadow boundary, both illumination and gradient based methods cannot well handle the boundary problems, especially in the presence of complex texture or color distortion.

Recently, deep neural networks are widely introduced for shadow removal through analyzing and learning the mapping relation between shadow image and the corresponding shadow-free image. Hu {\em et al.} \cite{Hu2018Direction} used multiple convolutional neural networks to learn image features for shadow detection 
and remove shadows in the image. Qu {\em et al.} \cite{Qu2017DeshadowNet} proposed an end-to-end DeshadowNet to recover illumination in shadow regions. 
Wang {\em et al.} \cite{Wang2017Stacked} proposed a stacked conditional generative adversarial network (ST-CGAN) for image shadow removing. Sidorov \cite{Sidorov2018Conditional} proposed an end-to-end architecture named AngularGAN oriented specifically to the color constancy task, without estimating illumniation color or illumination color map. 
Wei {\em et al.} \cite{WeiCGF2019} proposed a two-stage generative adversarial network for shadow inpainting and removal with slice convolutions.
Ding {\em et al.} \cite{Ding2019ICCV} proposed an attentive recurrent generative adversarial network (ARGAN) to detect and remove shadow with multiple steps.
Different from existing methods, our proposed RIS-GAN makes full use of the explored negative residual image and the inverse illumination map for generating more accurate shadow-removal results.

\section{Approach}
We explore the residual and illumination between shadow images and shadow-free images via Generative Adversarial Networks (GANs) \cite{goodfellow2014generative} due to the ability of GAN in style transfer and details recovery \cite{li2016precomputed}. The intuition behind is that the residual and illumination explored can provide informative additional details and insights for shadow 
removal.

The proposed framework RIS-GAN for shadow 
removal with multiple GANs is illustrated in Figure \ref{fig:pipeline}. Given an input shadow image $I$, three encoder-decoder structures are applied to generate residual image $I_{res}$, intermediate shadow-removal image $I_{imd}$, and inverse illumination map $S_{inv}$. With the element-wise addition with the input shadow image and the residual image, we are able to get an indirect shadow-removal image $I_{rem}^1$. With the element-wise production with the input shadow image and the inverse illumination, we are able to get another indirect shadow-removal image $I_{rem}^2$. 
We can apply another encoder-decoder structure to refine the coarse shadow-removal image $I_{coarse}$ and the two indirect shadow-removal images $I_{rem}^1$ and $I_{rem}^2$ to produce a fine shadow-removal image $I_{fine}$.

Our RIS-GAN is composed of four generators in the same encoder-decoder structure and three discriminators. The four generators are residual generator, removal generator, illumination generator, detection generator, and refinement generator, denoted as $G_{res}$, $G_{rem}$, $G_{illum}$, 
and $G_{ref}$, respectively, for generating a negative residual image, a coarse shadow-removal image, an inverse illumination map, 
and a fine shadow-removal image. The three discriminators, $D_{res}$, $D_{illum}$ and $D_{ref}$, share the same architecture and the same parameters to determine the generated residual images, inverse illumination maps, and final shadow-removal images to be real or fake, compared with the ground-truth residual images, inverse illumination maps, and the shadow-free images. For the readers¡¯ convenience, we summarize the relations between the above mentioned notations as:
\begin{equation}
I_{res} = G_{res}(I), S_{inv} = G_{illum}(I)
\end{equation}
\begin{equation}
    I_{rem}^1 = I\oplus I_{res}, I_{rem}^2 = I\otimes S_{inv}
\end{equation}
\begin{equation}
I_{coarse} = G_{rem}(I)
\end{equation}
\begin{equation}
    I_{fine} = G_{ref}(I_{coarse}, I_{rem}^1, I_{rem}^2)
\end{equation}


Our network takes the shadow image as input and outputs the negative residual images, inverse illumination maps, and fine shadow-removal images in an end-to-end manner. The alternative training between generators and discriminators ensures the good quality of the prediction results.

In the following, we are going to describe the details of our generators, discriminators, loss functions, as well as the implementation details.

\subsection{Encoder-Decoder Generators}
In principle, any encoder-decoder structures can be used in our RIS-GAN framework. In this paper, we don't want to design any particular shadow-aware components in the framework, and just adopt the DenseUNet architecture \cite{Bharath2018Single} as the implementation of each encoder-decoder generator. DenseUNet consists of a contracting path to capture context and a symmetric expanding path to upsample. Different with the conventional UNet architecture, DensUNet adds Dense Blocks in the network, which concatenate every layer¡¯s output with its input, and feed it to the next layer. This enhances information and gradient flow in our four encoder-decoder generators:
\begin{itemize}
    \item Residual Generator $G_{res}$ is to get a residual image that is close to the ground-truth residual image $I_{res}^{gt}$ obtained between shadow image and the corresponding shadow-free image $I^{gt}$, {\em i.e.}, $I_{res}^{gt} = I^{gt} - I$.
    \item Removal Generator $G_{rem}$ is to produce a coarse shadow-removal image $I_{coarse}$.
    \item Illumination Generator $G_{illum}$ is to estimate the inverse illumination map in the shadow image. Note that the ground-truth inverse illumination map is calculated based on the Retinex-based image enhancement methods~\cite{fu2016weighted,guo2016lime,wang2019underexposed}, {\em i.e.}, $S_{inv}^{gt} = I^{gt}\ast I^{-1}$, where $I^{gt}$ can be considered as a reflectance image, and $I$ is the observed image.
    \item Refinement Generator $G_{ref}$ is to refine the current intermediate shadow-removal image and two indirect shadow-removal images with the explored residual and illumination to formulate the final shadow-removal image.
\end{itemize}

\begin{figure}[ht!]
  \centering
  \includegraphics[width=0.185\linewidth]{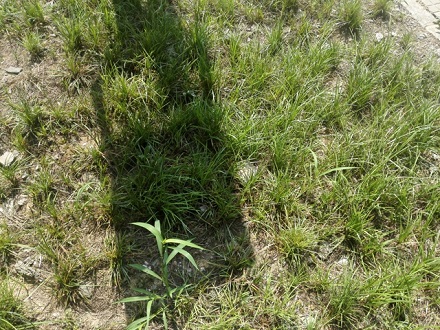}
  \includegraphics[width=0.185\linewidth]{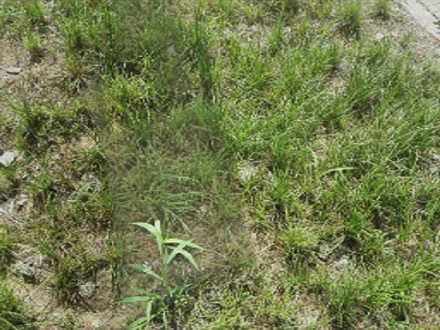}
  \includegraphics[width=0.185\linewidth]{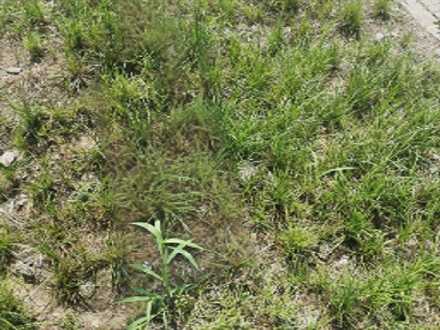}
  \includegraphics[width=0.185\linewidth]{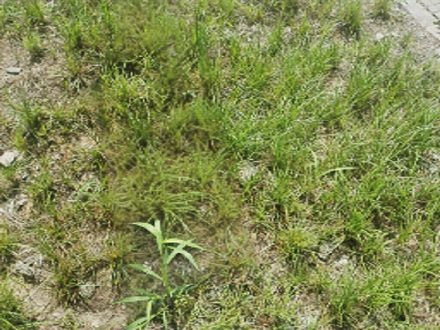}
  \includegraphics[width=0.185\linewidth]{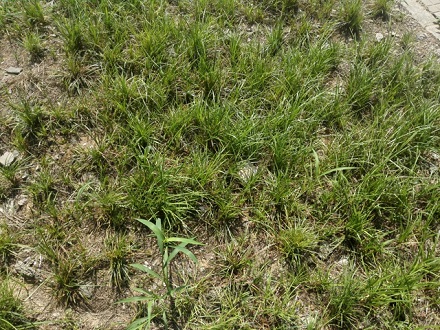}\\
  \vspace{2pt}
  \includegraphics[width=0.185\linewidth]{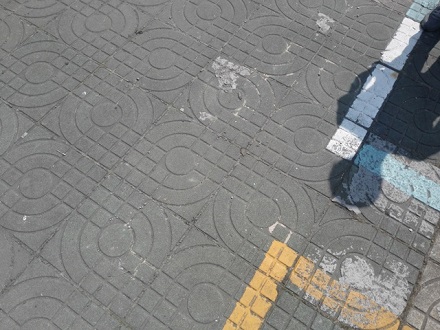}
  \includegraphics[width=0.185\linewidth]{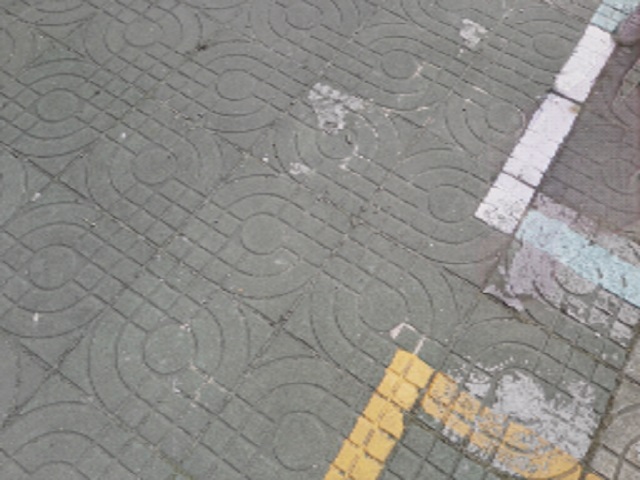}
  \includegraphics[width=0.185\linewidth]{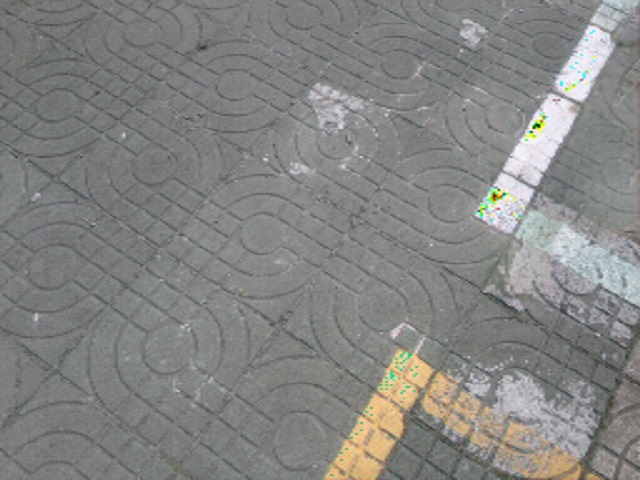}
  \includegraphics[width=0.185\linewidth]{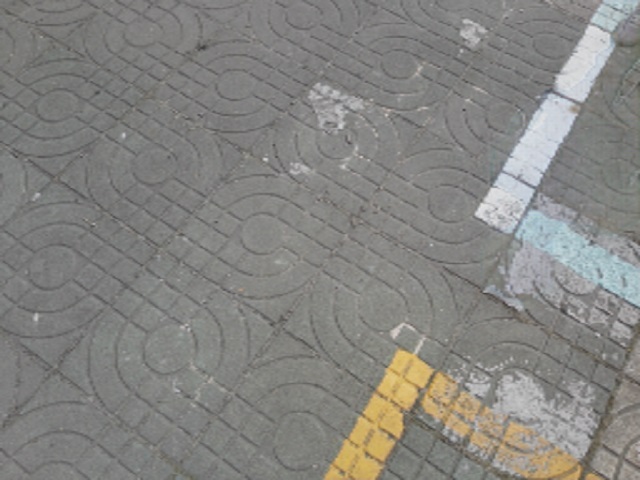}
  \includegraphics[width=0.185\linewidth]{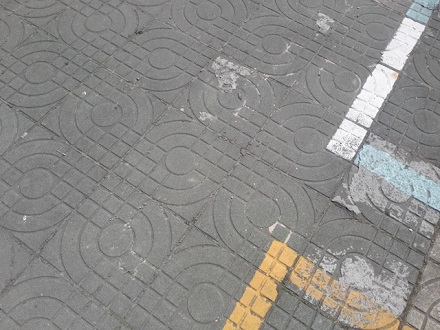}
  \vspace{-0.3cm}
  \caption{The visualization of residual and illumination for shadow removal. From left to right are the shadow images $I$, the indirect shadow-removal image $I_{rem}^1$ by residual, the indirect shadow-removal image $I_{rem}^2$ by illumination, the fine shadow-free image $I_{fine}^1$, and the ground-truth shadow-free image $I^{gt}$, respectively.}
\label{fig:example1_rem}
\end{figure}

To better understand our detector generator and refinement generators, we visualize some examples in 
Figure \ref{fig:example1_rem}. As we can observe, 
the indirect shadow-removal images obtained by residual and illumination have good quality and are complimentary to the intermediate shadow-removal image for further refinement to get the final shadow-removal image.

\subsection{Joint Discriminator}

The discriminator is a convolutional network, which is used to distinguish the predicted residual image, the final shadow-removal image, and the estimated illumination produced by the generators to be real or fake, compared with the corresponding ground truth. To make sure all the produced results are indistinguishable from the corresponding ground truths, we make use of a GAN with joint discriminator \cite{He2018Densely}. The joint discriminator is trained to learn a joint distribution to judge whether the produced results are real or fake.

\begin{figure}[ht!]
  \centering
  \includegraphics[width=0.95\linewidth]{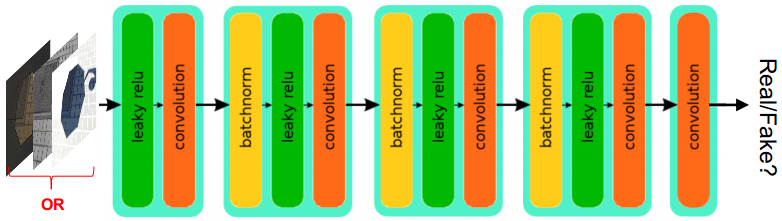}
  \vspace{-0.3cm}
  \caption{The architecture of the discriminator in our RIS-GAN. It consists of five convolution layers with batchnorm and leakly ReLU activations. For all these five convlution layers, all the kernel sizes are $4\times 4$; the strides are $4\times 4$ except the first convolution layer whose stride is $2\times 2$; and the number of output channels is: $64\rightarrow128\rightarrow256\rightarrow512\rightarrow1$. }
\label{fig:discriminator}
\end{figure}

Our discriminator consists of five convolution layers, each followed by a batch normalization and a Leaky ReLU activation function, and one fully connected layer. The last fully connected layer outputs the probability value that the input image (result produced by generator) is a real image. Figure \ref{fig:discriminator} gives details of the discriminator.


It is worth noting that we use the spectrum normalization method \cite{miyato2018spectral} to stabilize the training process of discriminator network, because spectral normalization is a simple and effective standardized method for limiting the optimization process of the discriminator in GAN, and it can make the whole generators perform better.


\subsection{Loss Functions}

To get a robust parametric mode, the loss functions that we use to optimize the proposed RIS-GAN has five components: shadow removal loss $\mathcal{L}_{rem}$, residual loss $\mathcal{L}_{res}$, illumination loss $\mathcal{L}_{illum}$, cross loss $\mathcal{L}_{cross}$ and adversarial loss $\mathcal{L}_{adv}$. The total loss $\mathcal{L}$ is can be written as
\begin{equation}
\mathcal{L}=\lambda_1\mathcal{L}_{res}+ \lambda_2\mathcal{L}_{rem} + \lambda_3\mathcal{L}_{illum} + \lambda_4\mathcal{L}_{cross}  + \mathcal{L}_{adv},
\label{eqn:overallloss}
\end{equation}
where $\lambda_1$, $\lambda_2$, $\lambda_3$, and $\lambda_4$ 
are hyperparameters.


\textbf{Shadow removal loss} is defined with visual-consistency loss and perceptual-consistency loss, {\em i.e.},
\begin{equation}
\mathcal{L}_{rem}=\mathcal{L}_{vis} +\beta_1\mathcal{L}_{percept},
\label{eq2}
\end{equation}
where $\beta$ is the weight parameter. $\mathcal{L}_{vis}$ is visual-consistency loss for removal generator which is calculated using L1-norm between the shadow removal result and the ground truth, and $\mathcal{L}_{percept}$ is perceptual-consistency loss aiming to preserve image structure. To specify,
\begin{equation}
\mathcal{L}_{vis}=||I^{gt} - I_{fine}||_1 + ||I^{gt} - I_{coarse}||_1.
\end{equation}
\begin{equation}
\begin{split}
    \mathcal{L}_{percept}&=||\text{VGG}(I^{gt}) - \text{VGG}(I_{fine}))||_2^2 \\
    &+ ||\text{VGG}(I^{gt}) - \text{VGG}(I_{coarse}))||_2^2.
\end{split}
\label{eq4}
\end{equation}
where $\text{VGG}(\cdot)$ is the feature extractor from the VGG19 model. 



\textbf{Residual loss} can be perceived as the obscured brightness in shadow regions, {\em i.e.}, 
\begin{equation}
\mathcal{L}_{res}=||I_{res}^{gt} - G_{res}(I)||_1.
\label{eq5}
\end{equation}

\textbf{Illumination loss} 
calculates L1-norm between the illumination result generated by $G_{illum}$ and the ground truth of inverse illumination map $S_{inv}^{gt}$. Then illumination loss for illumination branch can be denoted as:
\begin{equation}
\mathcal{L}_{illum}= ||S_{inv}^{gt} - G_{illum}(I)||_1.
\label{eq6}
\end{equation}

\textbf{Cross loss} is designed to ensure the consistency and correlation among residual, illumination and shadow information as
\begin{equation}
\mathcal{L}_{cross}=||I^{gt} - (G_{res}(I) \oplus I)||_1 + \beta_2 ||I^{gt} - (G_{illum}(I) \otimes I)||_1.
\label{eq7}
\end{equation}

\textbf{Adversarial loss} $\mathcal{L}_{adv}$ is the joint adversarial loss for the network, and is described as:
\begin{equation}
\begin{split}
\mathcal{L}_{adv}&=\mathbb{E}_{(I, I^{gt}, I_{res}^{gt}, S_{inv}^{gt})}[log(D_{ref}(I^{gt}))\\
&+log(1-D_{ref}(G_{ref}(G_{res}(I), G_{rem}(I), G_{illum}(I)))) \\
&+log(D_{res}(I_{res}^{gt})) + log(1-D_{res}(G_{res}(I))) \\
&+log(D_{illum}(S_{inv}^{gt})) + log(1-D_{illum}(G_{illum}(I)))]
\end{split}
\label{eq8}
\end{equation}
where $D_{res}$, $D_{ref}$, and $D_{illum}$ are the three discriminators.

Overall, our objective for the training task is solving a mini-max problem which aims to find a saddle point between generator and discriminator of our network.

\subsection{Implementation Details}
Our proposed method is implemented in Tensorflow in a computer with Intel(R) Xeon(R) Silver 4114 CPU@2.20GHz 192G RAM NVIDIA GeForce GTX 1080Ti. In our experiments, the input size of image is 256$\times$256. The learning rate value is set to 0.001. The parameters $\lambda_1$, $\lambda_2$, $\lambda_3$, $\lambda_4$, $\beta_1$ and $\beta_2$ are set to 10, 100, 1, 1, 0.1 and 0.2 in our experiments, respectively. The minibatch size is 2. The initial learning rate is set as 0.001. We use Momentum Optimizer to optimize our generator and use Adam Optimizer for the discriminator. We alternatively train our generator and discriminator for 10,000 epochs.

\begin{figure*}[ht!]
\centering
\includegraphics[width=0.10\textwidth]{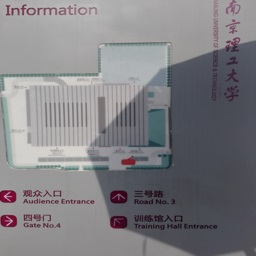}
\includegraphics[width=0.10\textwidth]{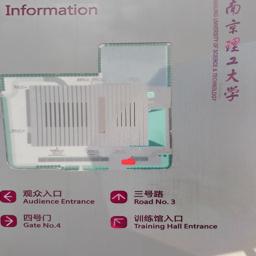}
\includegraphics[width=0.10\textwidth]{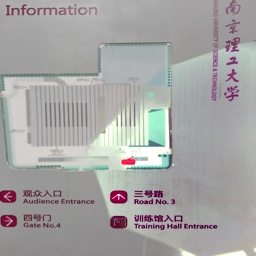}
\includegraphics[width=0.10\textwidth]{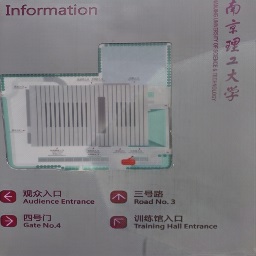}
\includegraphics[width=0.10\textwidth]{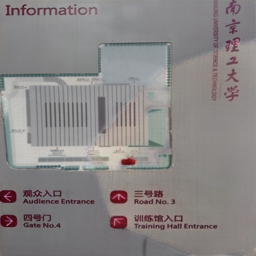}
\includegraphics[width=0.10\textwidth]{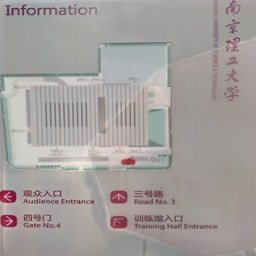}
\includegraphics[width=0.10\textwidth]{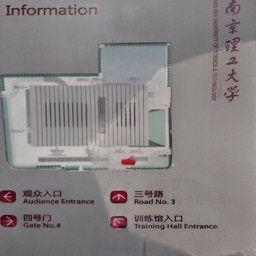}
\includegraphics[width=0.10\textwidth]{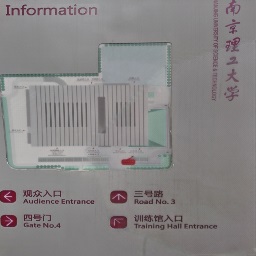}
\includegraphics[width=0.10\textwidth]{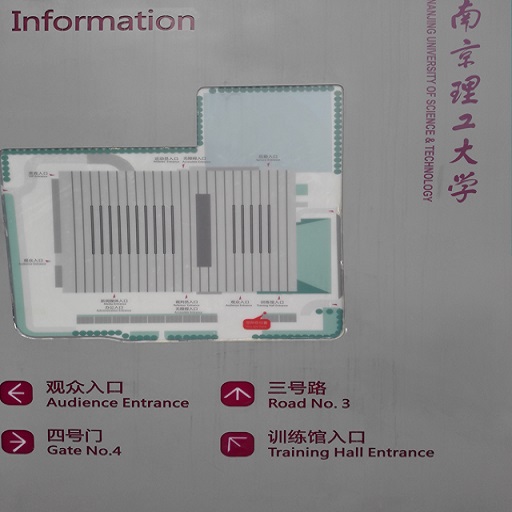}\\
 \vspace{-3pt}
  \subfigure[]{\includegraphics[width=0.10\textwidth]{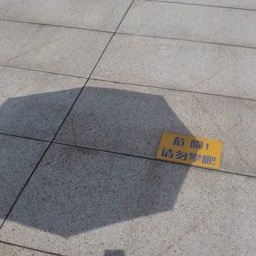}}
  \subfigure[]{\includegraphics[width=0.10\textwidth]{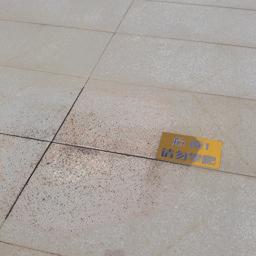}}
  \subfigure[]{\includegraphics[width=0.10\textwidth]{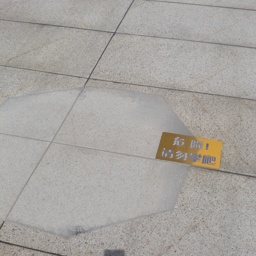}}
  \subfigure[]{\includegraphics[width=0.10\textwidth]{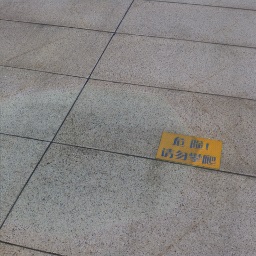}}
   \subfigure[]{\includegraphics[width=0.10\textwidth]{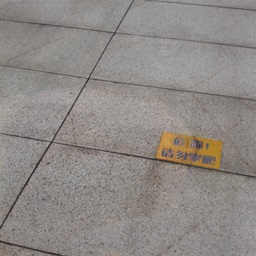}}
   \subfigure[]{\includegraphics[width=0.10\textwidth]{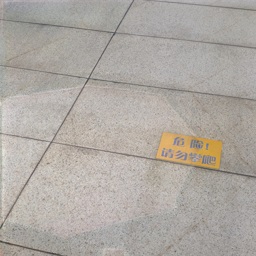}}
   \subfigure[]{\includegraphics[width=0.10\textwidth]{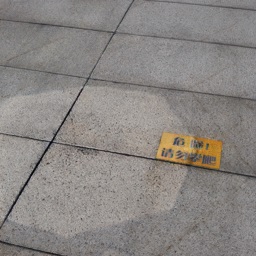}}
   \subfigure[]{\includegraphics[width=0.10\textwidth]{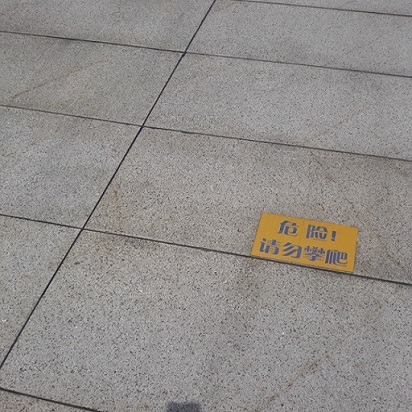}}
    \subfigure[]{\includegraphics[width=0.10\textwidth]{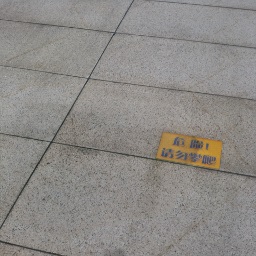}}
    \vspace{-0.5cm}
    \caption{Shadow removal results. From left to right are: input images (a); shadow-removal results of Guo (b), Zhang (c), DeshadowNet (d), ST-CGAN (e), DSC (f), AngularGAN (g), and our RIS-GAN (h); and the corresponding ground truth shadow-free images (i).
    }
    \label{fig:visualization_shadowremoval_dataset}
    \vspace{-0.3cm}
\end{figure*}

\section{Experiments}

To verify the effectiveness of our proposed RIS-GAN, we conduct various experiments on the SRD dataset \cite{Qu2017DeshadowNet} and the ISTD dataset \cite{Wang2017Stacked}. The SRD dataset has 408 pairs of shadow and shadow-free images publicly avaiable. The ISTD dataset contains 1870 image triplets of shadow image, shadow mask and shadow-free image. Such a dataset has 135 different simulated shadow environments and the scenes are very diverse. Both these two datasets contains various kinds of shadow scenes. In this paper, we use the 1330 pairs of shadow and shadow-free images from the ISTD dataset for training, and use the rest 540 pairs for testing. We also use the trained RIS-GAN model to evaluate on the SRD dataset.

Regarding the metrics, we use the root mean square error (RMSE) calculated in Lab space between the recovered shadow-removal image and the ground truth shadow-free image to evaluate the shadow removal performance. We also conduct a user study for comprehensive evaluation.


\begin{figure*}[ht!]
\centering
\includegraphics[width=0.115\textwidth]{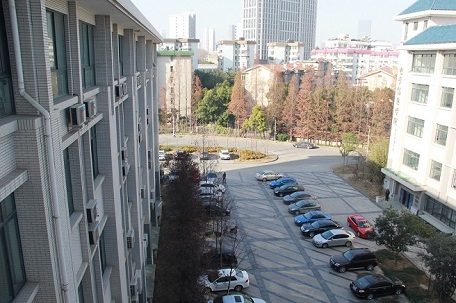}
\includegraphics[width=0.115\textwidth]{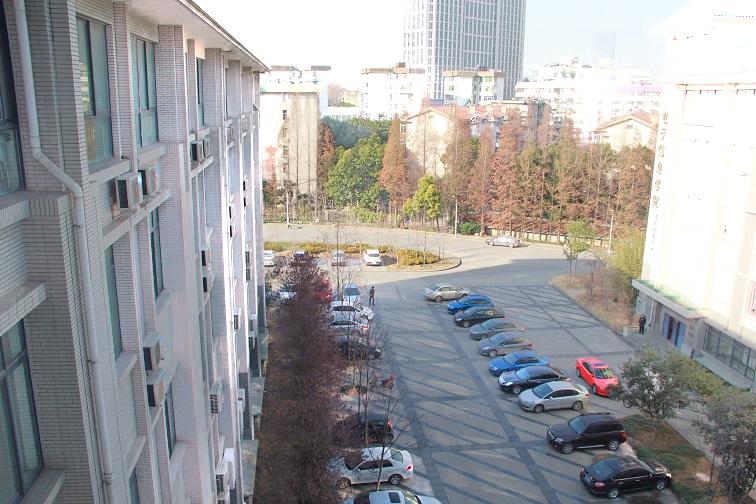}
\includegraphics[width=0.115\textwidth]{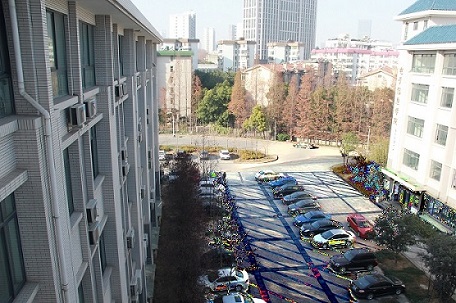}
\includegraphics[width=0.115\textwidth]{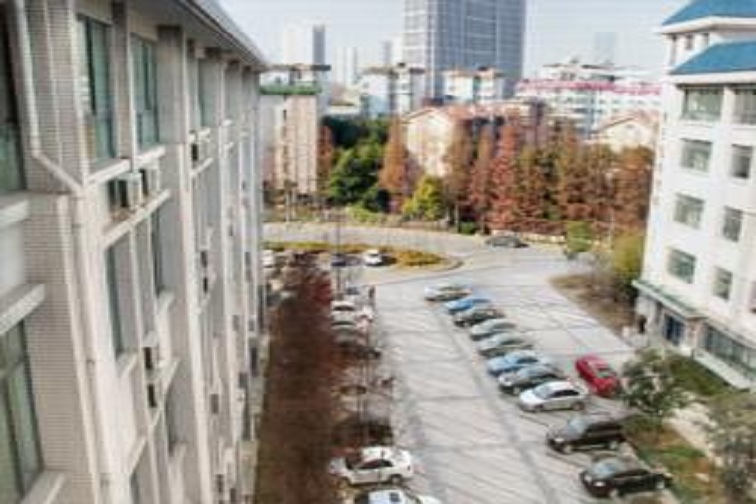}
\includegraphics[width=0.115\textwidth]{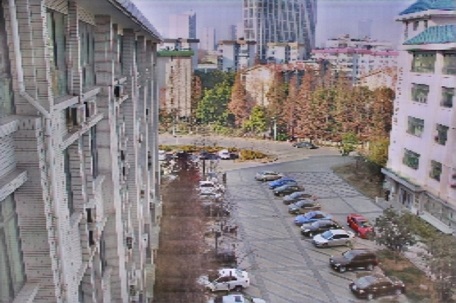}
\includegraphics[width=0.115\textwidth]{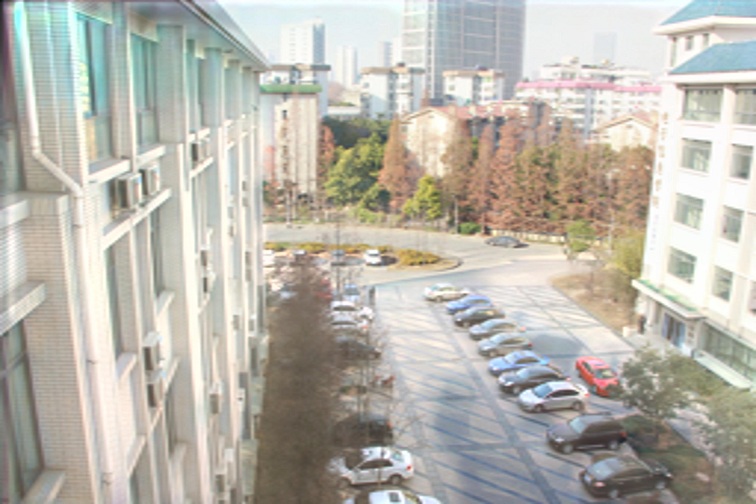}
\includegraphics[width=0.115\textwidth]{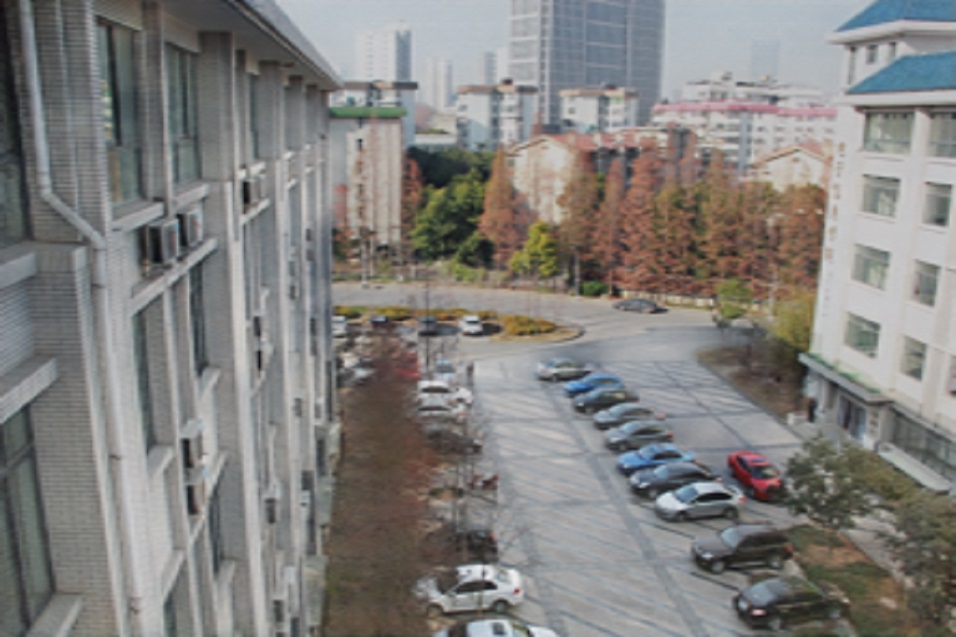}
\includegraphics[width=0.115\textwidth]{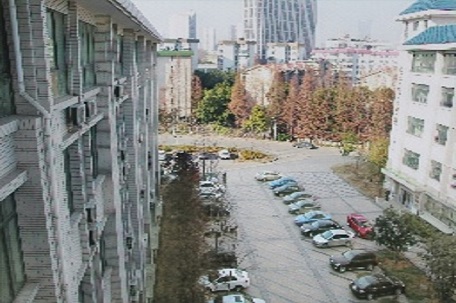}\\
 \vspace{2pt}
\includegraphics[width=0.115\textwidth]{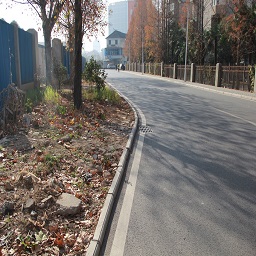}
\includegraphics[width=0.115\textwidth]{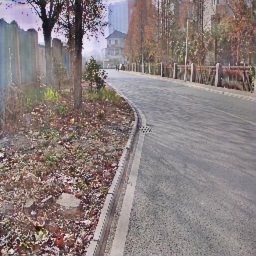}
\includegraphics[width=0.115\textwidth]{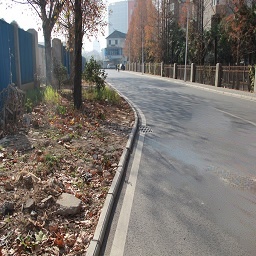}
\includegraphics[width=0.115\textwidth]{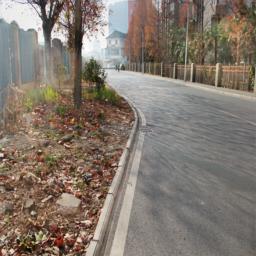}
\includegraphics[width=0.115\textwidth]{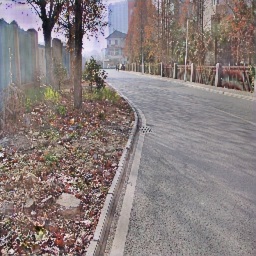}
\includegraphics[width=0.115\textwidth]{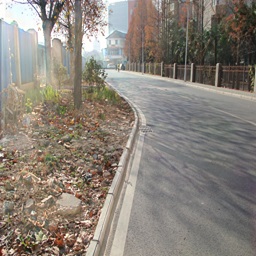}
\includegraphics[width=0.115\textwidth]{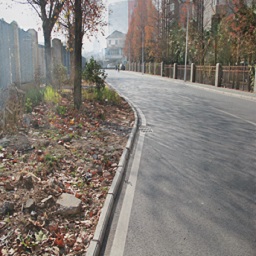}
\includegraphics[width=0.115\textwidth]{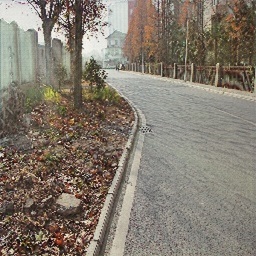}\\
 \vspace{-3pt}
\subfigure[]{\includegraphics[width=0.115\textwidth]{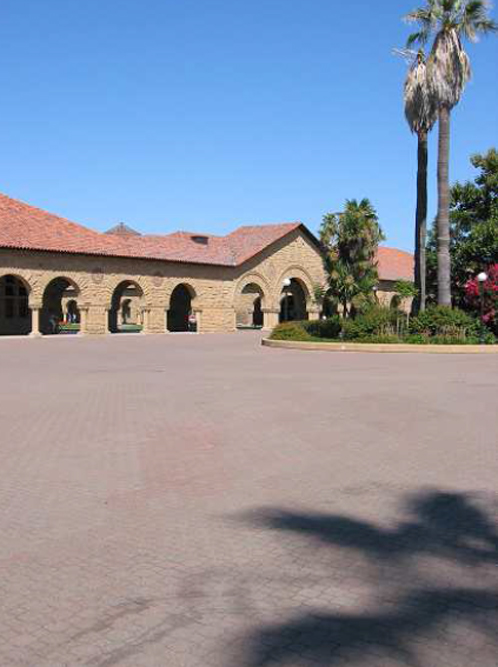}}
\subfigure[]{\includegraphics[width=0.115\textwidth]{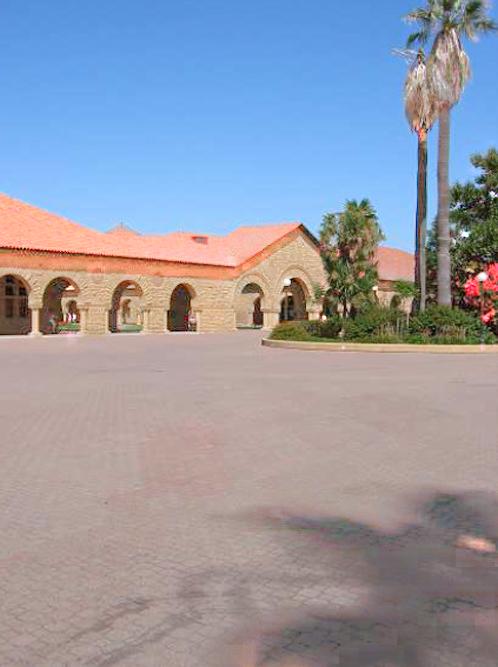}}
\subfigure[]{\includegraphics[width=0.115\textwidth]{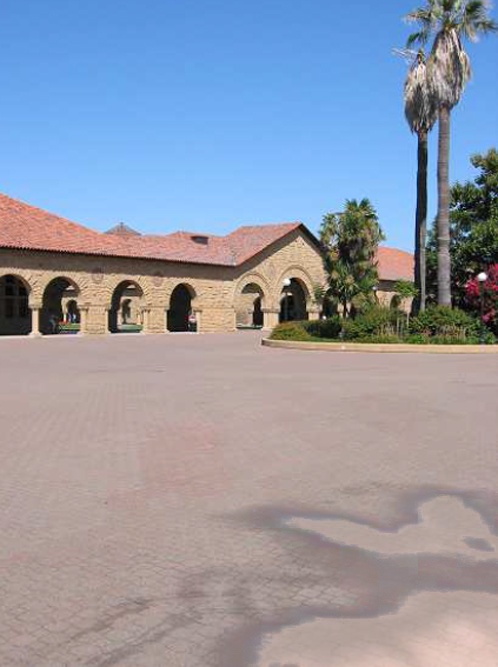}}
\subfigure[]{\includegraphics[width=0.115\textwidth]{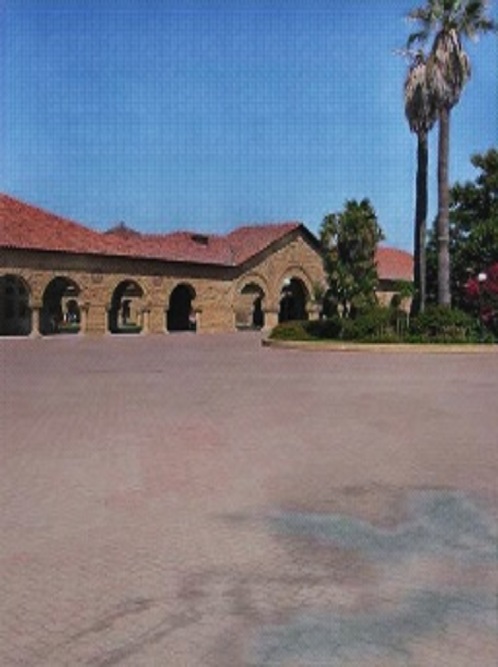}}
\subfigure[]{\includegraphics[width=0.115\textwidth]{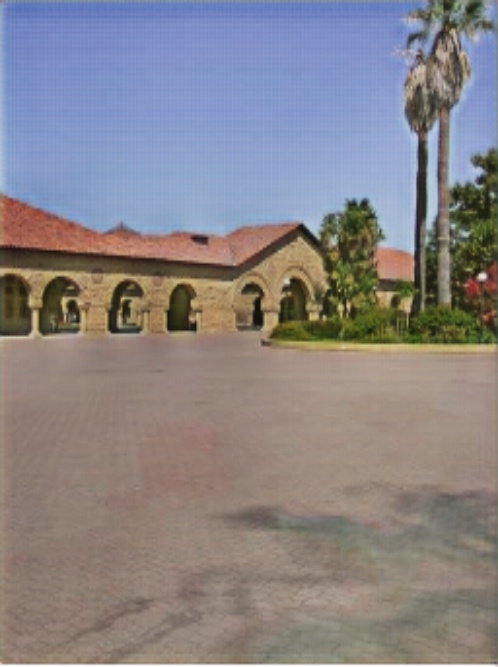}}
\subfigure[]{\includegraphics[width=0.115\textwidth]{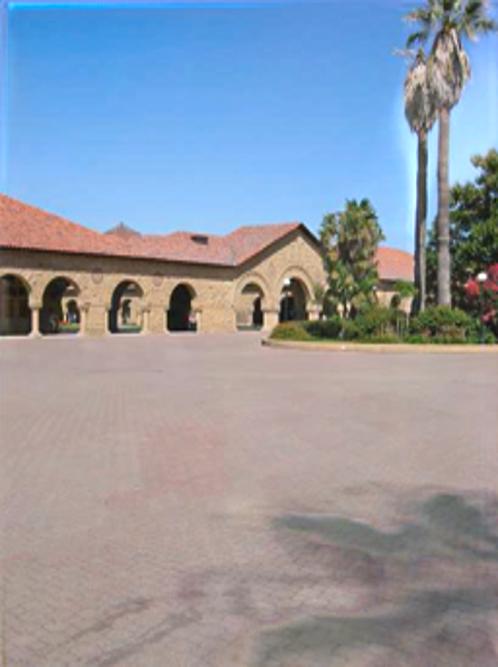}}
\subfigure[]{\includegraphics[width=0.115\textwidth]{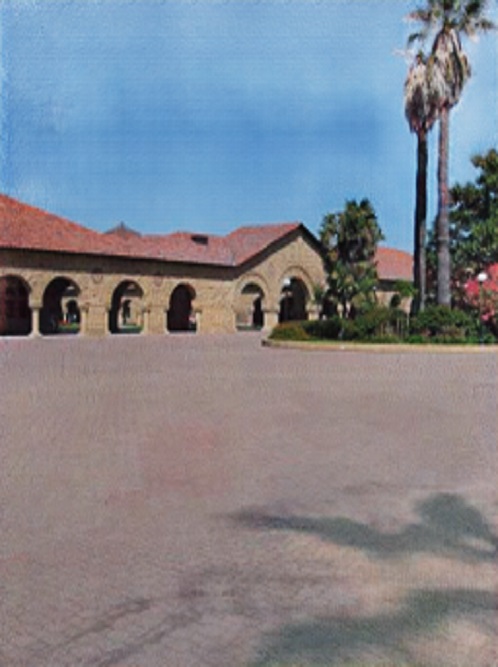}}
\subfigure[]{\includegraphics[width=0.115\textwidth]{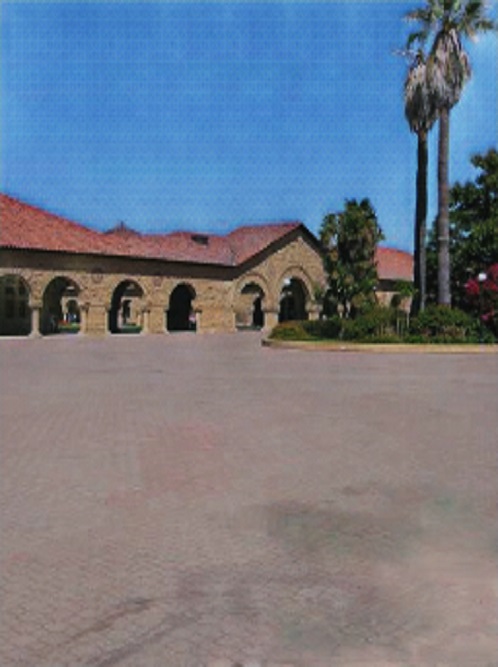}}
\vspace{-0.5cm}
\caption{Shadow removal results. From left to right are: input images (a); shadow-removal results of Guo (b), Zhang (c), DeshadowNet (d), ST-CGAN (e), DSC (f), and AngularGAN (g); and shadow-removal results of our RIS-GAN.
}
    \label{fig:visualization_internet}
    \vspace{-0.3cm}
\end{figure*}

\subsection{Comparison with State-of-the-arts}
We compare our RIS-GAN with the state of-the-art methods including the two traditional methods, {\em i.e.}, Guo \cite{Guo2011Single} and Zhang \cite{Zhang2015Shadow} and the recent learning-based methods, i.e., DeshadowNet \cite{Qu2017DeshadowNet}, DSC \cite{Hu2018Direction}, ST-CGAN \cite{Wang2017Stacked}, and AngularGAN ~\cite{Sidorov2018Conditional}. Note that shadow removal works on the pixel and recovers the value of the pixel, and therefore we add two frameworks, {\em i.e.}, Global/Local-GAN \cite{iizuka2017globally} for image inpainting and Pix2Pix-HD \cite{wang2018high} for image translation, as another two shadow removal baselines for solid validation. To make the fair comparison, we use the same training data with the same input size of images ($256\times256)$ to train all the learning-based methods on the same hardware.


We summarize the comparison results in Table \ref{tab:shadow_removal_srd} and Table \ref{tab:shadow_removal_istd}.
From the table, we can observe that among all the competing methods, our proposed RIS-GAN achieves the best RMSE values in shadow regions, non-shadow regions, and the entire images on the two datasets, although we have no particular shadow-aware components designed in our generators. This suggests that the recovered shade-removal images obtained by our RIS-GAN is much closer to the corresponding ground-truth shadow-free images. As the main difference between our RIS-GAN and the state-of-the-art deep learning, exploring residual and illuminations demonstrates the great advantages in the task of shadow removal.


\begin{table}[ht!]
   \small
   \centering
   \caption{Quantitative comparison results of shadow removal on the SRD dataset using the metric RMSE (the smaller, the better). S, N, and A represent shadow regions, non-shadow region, and the entire image, respectively.}
   \label{tab:shadow_removal_srd}  
   \begin{tabular}{c|c|ccc}
   \hline
      Methods & Venue/Year &  S  & N & A \\
      \hline
      Guo    & CVPR/2011 &   31.06 & 6.47   &12.60\\
      Zhang  & TIP/2015 & 9.50  & 6.90  &7.24  \\
      \hline
      Global/Local-GAN & TOG/2017 & 19.56 & 8.17 & 16.33 \\
      Pix2Pix-HD & CVPR/2018 & 17.33 & 7.79 & 12.58 \\
      \hline
      Deshadow & CVPR/2017 &  17.96 & 6.53  &8.47 \\
      ST-CGAN  & CVPR/2018 &   18.64  & 6.37  &8.23 \\
      DSC      & CVPR/2018 &  11.31 &  6.72 & 7.83  \\
      AgularGAN& CVPRW/2019 & 17.63 &  7.83 & 15.97 \\
      \hline
      RIS-GAN  & AAAI/2020 & {\bf 8.22} & {\bf 6.05} & {\bf 6.78}   \\
      \hline
   \end{tabular}
   \vspace{-0.3cm}
\end{table}

\begin{table}[ht!]
   \small
   \centering
   \caption{Quantitative comparison results of shadow removal on the ISTD dataset in term of RMSE. 
   }
   \label{tab:shadow_removal_istd}  
   \begin{tabular}{c|c|ccc}
   \hline
      Methods & Venue/Year &  S  & N & A \\
      \hline
        Guo      &  CVPR/2011    & 18.95 & 7.46 & 9.30   \\
        Zhang  & TIP/2015  &9.77 & 7.12  &8.16  \\
      \hline
        Global/Local-GAN & TOG/2017 & 13.46 & 7.67 & 8.82 \\
        Pix2Pix-HD & CVPR/2018 & 10.63 & 6.73 & 7.37 \\
      \hline
        Deshadow & CVPR/2017  &12.76  &7.19  & 7.83  \\
        ST-CGAN    & CVPR/2018  &10.31&6.92   & 7.46  \\
        DSC      &  CVPR/2018  & 9.22&6.50&7.10  \\
        AngularGAN& CVPRW/2019  & 9.78&7.67&8.16  \\
       \hline
        RIS-GAN  & AAAI/2020 & {\bf 8.99} & {\bf 6.33}   & {\bf 6.95}   \\
        \hline
  \end{tabular}
  \vspace{-0.1cm}
\end{table}


To further explain the outperformance of our proposed RIS-GAN, we provides some visualization results in Figure \ref{fig:visualization_shadowremoval_dataset} covering the traditional methods and the learning-based methods for shadow removal. As we can see in Figure \ref{fig:visualization_shadowremoval_dataset}(b), Guo can recover illumination in shadow regions and may produce unnatural shadow removal results especially for images with different textures in the shadow regions. Zhang cannot well handle the illumination change in shadow boundaries so that the recovered shadow-removal images have boundary problem, such as color distortion or texture loss, as shown in Figure \ref{fig:visualization_shadowremoval_dataset}(c). 
Compared with these two traditional methods, our proposed RIS-GAN not only effectively recovers illumination in shadow regions, but also reconstruct the illumination and texture in shadow boundaries, as shown in Figure \ref{fig:visualization_shadowremoval_dataset}(h).

As for the recent deep learning methods, DeshadowNet, ST-CGAN, and DSC deal with the images in aspect of color space, without considering the aspect of illumination. This may lead to unsatisfied shadow-removal results like color distortion or incomplete shadow removal, as shown in Figure \ref{fig:visualization_shadowremoval_dataset} (d-f). AngularGAN introduces an angle loss which is defined based on the consideration of illumination, which makes the illumination of non-shadow regions very close to the corresponding ground-truth shadow-free images.
It is worth mentioning here that the illumination is not well incorporated in the recovery of shadow-removal images, which causes some inconsistency between shadow-region and non-shadow region, as seen in Figure \ref{fig:visualization_shadowremoval_dataset} (g). In contrast, taking residual negative residual images and the inverse illumination maps into consideration, our proposed RIS-GAN can effectively remove shadows and produce good result for both simple and complex scene image. The recovered illumination in shadow regions is consistent with surrounding environment and the texture details in shadow regions are well preserved, as shown in the Figure \ref{fig:visualization_shadowremoval_dataset}(h).

To further verify the robustness and the potential of our proposed RIS-GAN in the complicated scenes, we collect a few shadow images in the real-world life to run the experiments and report in Figure \ref{fig:visualization_internet}. Aparently, the shadow-removal images recovered by our proposed RIS-GAN look more realistic, with little artifacts. This observation demonstrates the robustness of our RIS-GAN for complex scenes.

\textbf{User study}
We conduct a user-study with 100 random volunteers to evaluate the visual performance of our proposed RIS-GAN and some other shadow removal methods. We prepare 300 sets of images. Each set contains five shadow removal results by using methods of our RIS-GAN, DeshodowNet, ST-CGAN, DSC, and AgularGAN, respectively. For each volunteer, we randomly show them twenty sets images to choose which shadow removal image is the most natural in each set. Then there will be 2000 select results. Counting all the results, we find that 29.65\% of shadow-removal images generated by our RIS-GAN are chosen as the most natural shadow removal result, while 14.85\%, 19.55\%, 20.35\% and 15.60\% of shadow removal results are chosen by Deshodow, ST-CGAN, DSC, and AgularGAN,  respectively.

\begin{figure*}[ht!]
 \centering
 \includegraphics[width=0.10\textwidth]{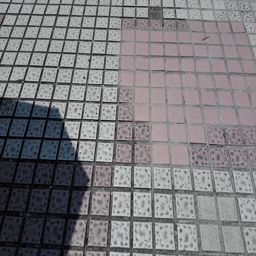}
 \includegraphics[width=0.10\textwidth]{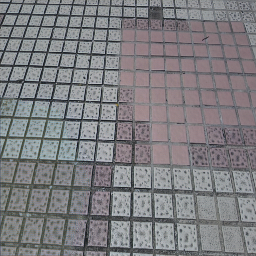}
 \includegraphics[width=0.10\textwidth]{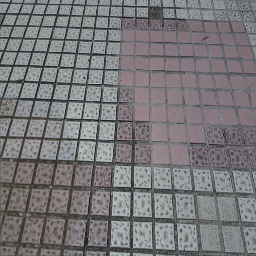}
 \includegraphics[width=0.10\textwidth]{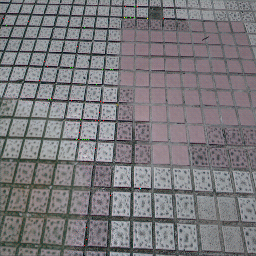}
 \includegraphics[width=0.10\textwidth]{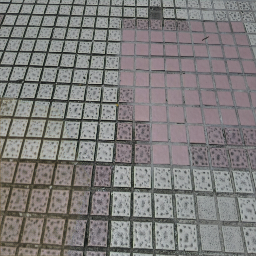}
 \includegraphics[width=0.10\textwidth]{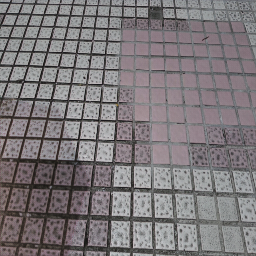}
 \includegraphics[width=0.10\textwidth]{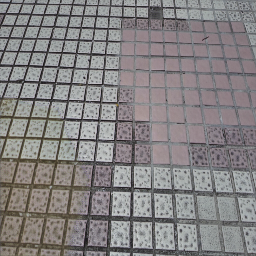}
 \includegraphics[width=0.10\textwidth]{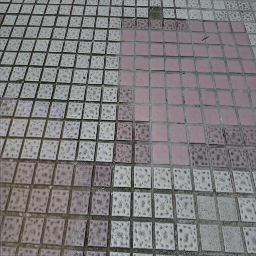}
 \includegraphics[width=0.10\textwidth]{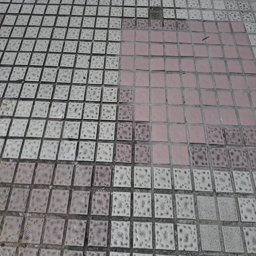}\\
 \vspace{-3pt}
  \subfigure[]{\includegraphics[width=0.10\textwidth]{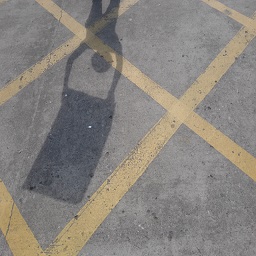}}
  \subfigure[]{\includegraphics[width=0.10\textwidth]{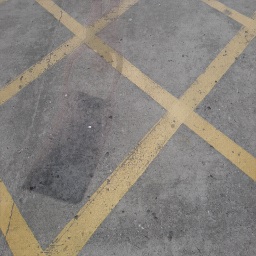}}
 \subfigure[]{\includegraphics[width=0.10\textwidth]{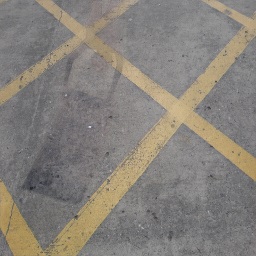}}
   \subfigure[]{\includegraphics[width=0.10\textwidth]{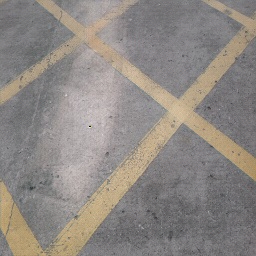}}
   \subfigure[]{\includegraphics[width=0.10\textwidth]{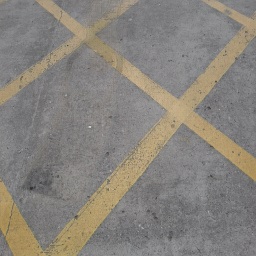}}
   \subfigure[]{\includegraphics[width=0.10\textwidth]{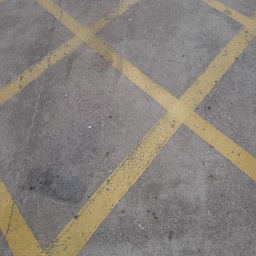}}
    \subfigure[]{\includegraphics[width=0.10\textwidth]{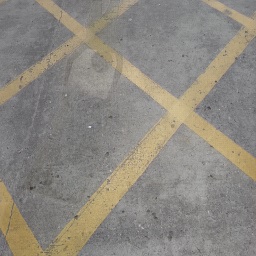}}
    \subfigure[]{\includegraphics[width=0.10\textwidth]{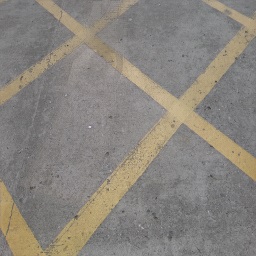}}
   \subfigure[]{\includegraphics[width=0.10\textwidth]{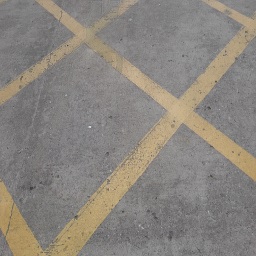}}
   \vspace{-0.4cm}
    \caption{Shadow removal results. From left to right are: input images (a); shadow-removal results of R-GAN (b), S-GAN (c), I-GAN (d),  RS-GAN (e), IS-GAN (f), $\text{RIS-GAN}_1$ (g), $\text{RIS-GAN}_2$ (h), and RIS-GAN (i), respectively.}
    \label{fig:ablation_study}
    \vspace{-0.3cm}
\end{figure*}

\subsection{Ablation Study}
To further evaluate some components of our proposed RIS-GAN, we design a series of variants as follows:
\begin{itemize}
    \item BASE: take the input shadow images as the shadow-removal result.
    \item R-GAN: use $G_{res}$ only and take $I_{rem}^1$ as the shadow-removal result.
    \item I-GAN: use $G_{illum}$ only and take $I_{rem}^2$ as the shadow-removal result.
    \item S-GAN: use $G_{rem}$ only and take $I_{coarse}$ as the shadow-removal result.
    \item RS-GAN: remove $G_{illum}$, and $G_{ref}$ takes $I_{res}$ and $I_{coarse}$ to get the fine shadow-removal image.
    \item IS-GAN: remove $G_{res}$, and $G_{ref}$ takes $S_{inv}$ and $I_{coarse}$ to get the fine shadow-removal image.
    \item $\text{RIS-GAN}_1$: remove $\mathcal{L}_{adv}$ from Equation~\ref{eqn:overallloss}.
    \item $\text{RIS-GAN}_2$: remove $\mathcal{L}_{cross}$ from Equation~\ref{eqn:overallloss}. 
\end{itemize}

We train the above seven GAN variants on the same training data and evaluate the shadow-removal results on both the SRD dataset and the ISTD dataset. The results are summarized in Table \ref{tab:ablation_study}, from which we can observe: (1) all the GAN variants can recover shadow-light or shadow-free in the shadow regions when compared with BASE; (2) the negative residual image from the residual generator and the inverse illumination map can help improve the performance of the shadow-removal refinement, and the combination leads to the best performance; and (3) the loss functions $\mathcal{L}_{adv}$ and $\mathcal{L}_{cross}$ are necessary to ensure the high-quality shadow-removal results, this clearly demonstrates the advantage of the joint discriminator and cross learning among three outputs. We also provide the visualization in Figure \ref{fig:ablation_study}, from which we can clearly see that our RIS-GAN recovers the best details of the shadow-removal regions and looks more realistic.
\begin{table}[ht!]
   \footnotesize
   \centering
   \caption{Quantitative shadow-removal results of ablation study on the SRD and ISTD datasets in term of RMSE.}
   \label{tab:ablation_study}  
   \begin{tabular}{c|ccc|ccc}
      \hline
      \multirow{2}{*}{Methods} & \multicolumn{3}{c|}{SRD} & \multicolumn{3}{c}{ISTD} \\
      & S  & N & A &  S  & N & A \\
      \hline
      BASE &   35.74  & 8.88 & 15.14 & 35.74  & 8.88 & 15.14  \\
      \hline
      R-GAN  & 11.41 & 7.33 &8.37 &12.09&7.08   &8.24   \\
      S-GAN  & 12.06 & 7.65 &8.85 &16.98&9.71   &11.27   \\
      I-GAN  & 14.82 & 8.54 &11.55  &9.02&  12.10 &15.31   \\
      RS-GAN & 10.33 &  6.44&7.35 &9.43&6.26   & 7.03  \\
      IS-GAN  & 9.57 & 6.32 &7.16 &10.16&6.37   &7.20   \\
      $\text{RIS-GAN}_1$& 9.37& 6.64 &7.32 &9.17&7.16&7.59 \\
      $\text{RIS-GAN}_2$& 9.51& 6.87 & 7.27 &11.01&8.98 &7.91  \\
      RIS-GAN  & {\bf 8.22} & {\bf 6.05} & {\bf 6.78} & {\bf 8.99} & {\bf 6.33}   & {\bf 6.95}   \\
      \hline
\end{tabular}
\vspace{-0.3cm}
\end{table}

\begin{figure}[ht!]
  \centering
  \includegraphics[width=0.19\linewidth]{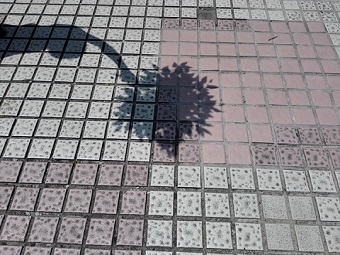}
  \includegraphics[width=0.19\linewidth]{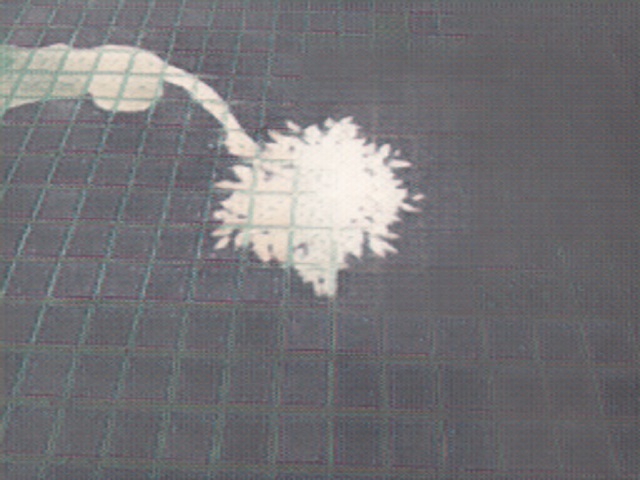}
  \includegraphics[width=0.19\linewidth]{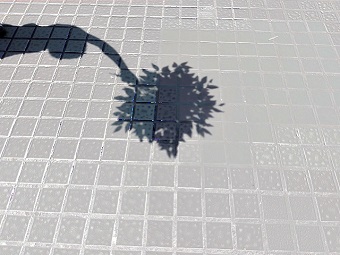}
  \includegraphics[width=0.19\linewidth]{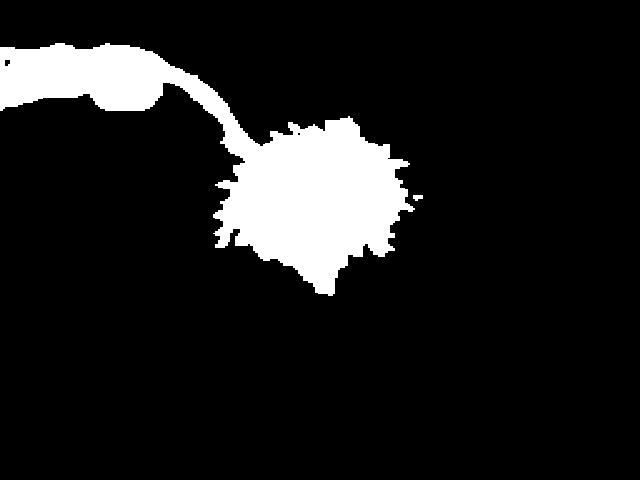}
  \includegraphics[width=0.19\linewidth]{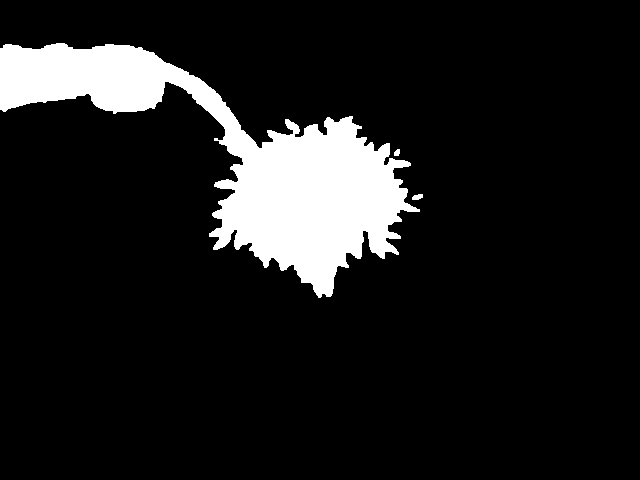}\\
  \vspace{2pt}
  {\includegraphics[width=0.19\linewidth]{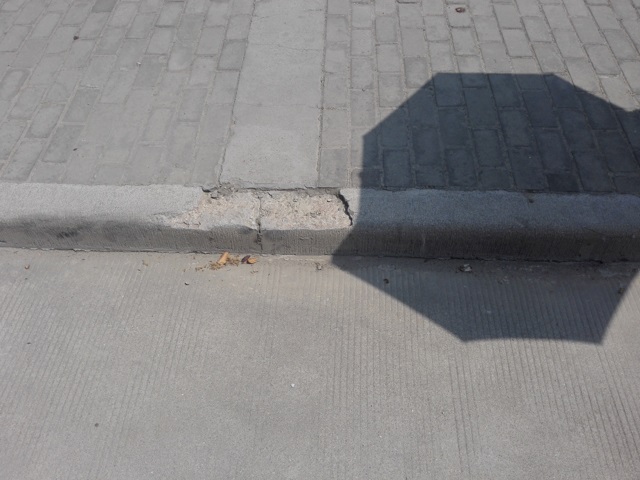}}   
  {\includegraphics[width=0.19\linewidth]{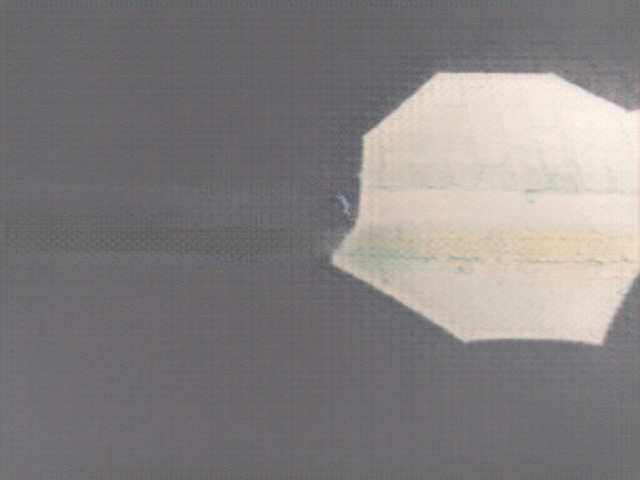}}   
  {\includegraphics[width=0.19\linewidth]{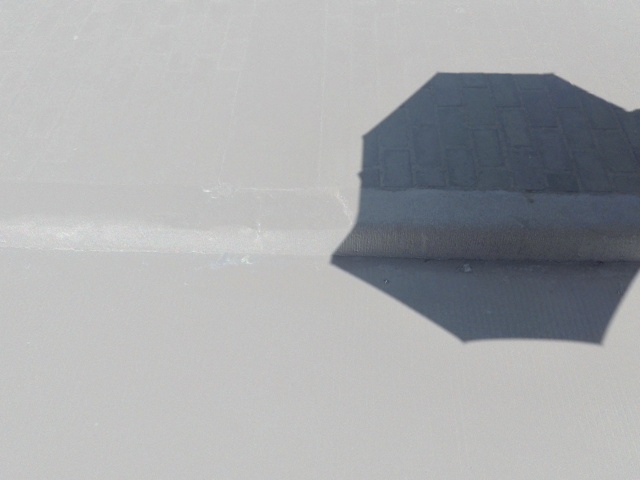}}   
  {\includegraphics[width=0.19\linewidth]{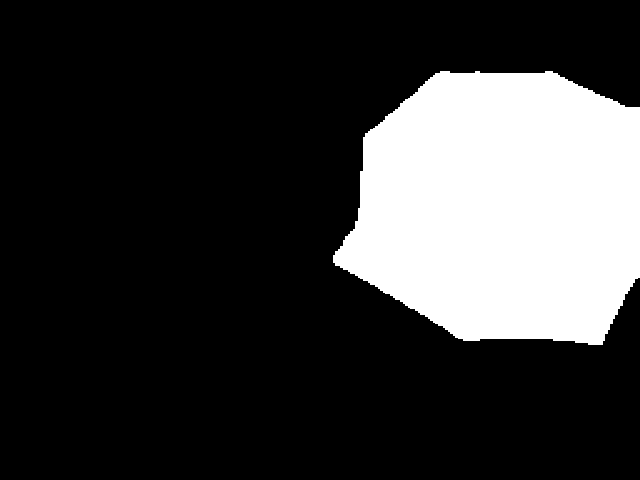}}
  {\includegraphics[width=0.19\linewidth]{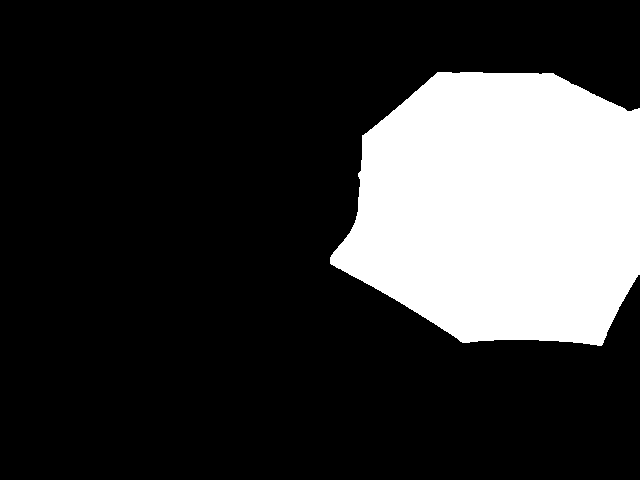}}
  \vspace{-0.3cm}
  \caption{The visualization of detection results. From left to right are input images, negative residual images, inverse illumination maps, prediction shadow masks based on the explored negative residual images and inverse illumination maps, and ground-truth shadow masks, respectively.}
  \label{fig:vis_det}
  \vspace{-0.3cm}
\end{figure}
\subsection{Discussion}
To better explore the potential of our proposed RIS-GAN, we also visualize the shadow detection masks, and extend the current approach for video shadow removal.

\textbf{Shadow detection}. Although we focus on shadow removal rather than detection, our RIS-GAN also can get the shadow detection masks based on the negative residual images and the inverse illumination maps. Figure \ref{fig:vis_det} shows the promising detection results. We observe that both the generated negative residual images and inverse illumination maps effectively distinguish the shadow and non-shadow regions well.

\begin{figure}[ht!]
  \centering
    \begin{minipage}[b]{0.115\linewidth}
    \includegraphics[width=1\linewidth]{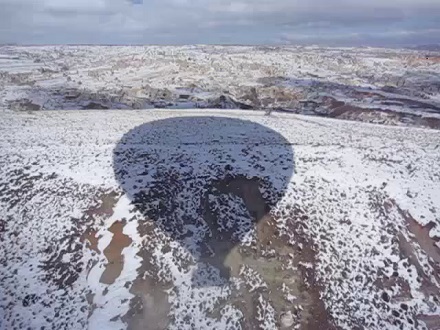}\vspace{1pt}
    \includegraphics[width=1\linewidth]{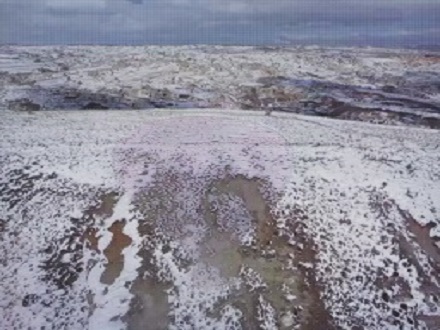}\vspace{1pt}
    \end{minipage}
  \hspace{0pt}
    \begin{minipage}[b]{0.115\linewidth}
    \includegraphics[width=1\linewidth]{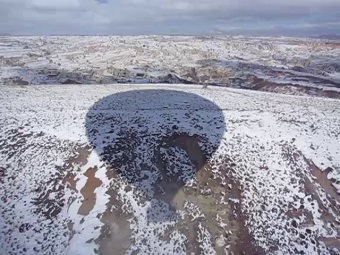}\vspace{1pt}
    \includegraphics[width=1\linewidth]{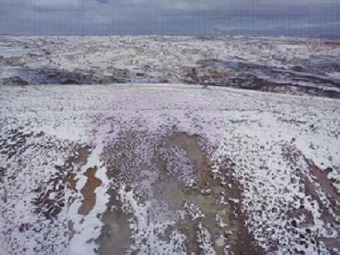}\vspace{1pt}
  \end{minipage}
  \hspace{0pt}
    \begin{minipage}[b]{0.115\linewidth}
    \includegraphics[width=1\linewidth]{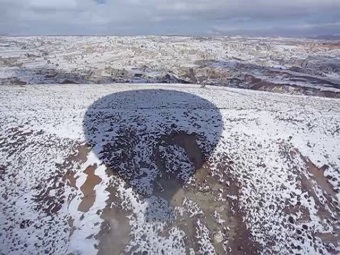}\vspace{1pt}
    \includegraphics[width=1\linewidth]{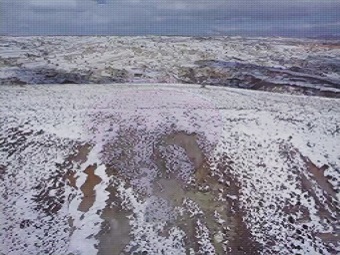}\vspace{1pt}
  \end{minipage}
  \hspace{0pt}
    \begin{minipage}[b]{0.115\linewidth}
    \includegraphics[width=1\linewidth]{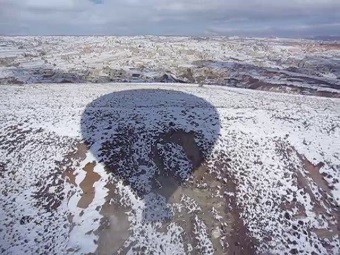}\vspace{1pt}
    \includegraphics[width=1\linewidth]{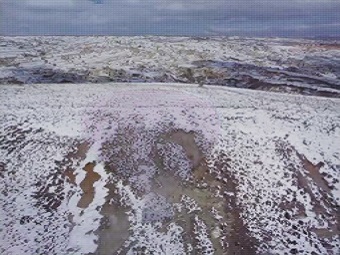}\vspace{1pt}
  \end{minipage}
  \hspace{0pt}
    \begin{minipage}[b]{0.115\linewidth}
    \includegraphics[width=1\linewidth]{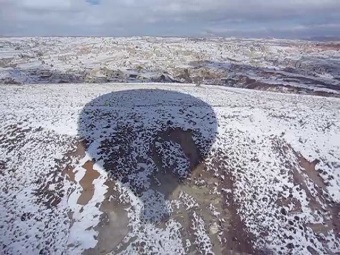}\vspace{1pt}
    \includegraphics[width=1\linewidth]{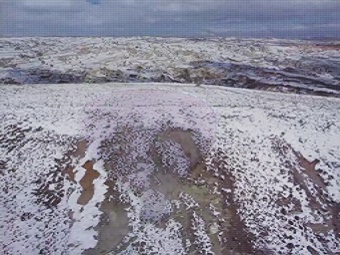}\vspace{1pt}
  \end{minipage}
  \hspace{0pt}
    \begin{minipage}[b]{0.115\linewidth}
    \includegraphics[width=1\linewidth]{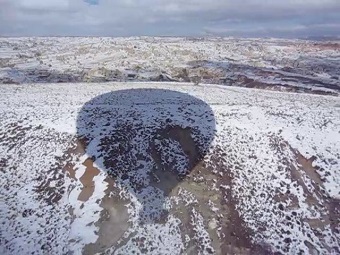}\vspace{1pt}
    \includegraphics[width=1\linewidth]{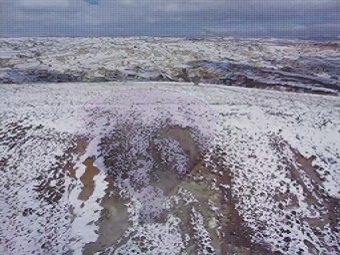}\vspace{1pt}
  \end{minipage}
  \hspace{0pt}
    \begin{minipage}[b]{0.115\linewidth}
    \includegraphics[width=1\linewidth]{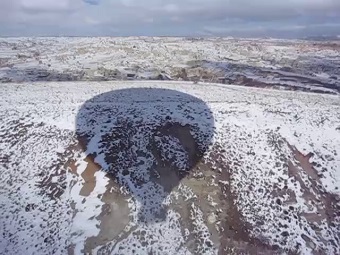}\vspace{1pt}
    \includegraphics[width=1\linewidth]{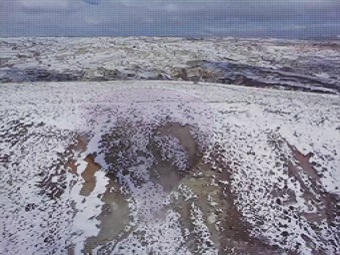}\vspace{1pt}
  \end{minipage}
  \hspace{0pt}
    \begin{minipage}[b]{0.115\linewidth}
    \includegraphics[width=1\linewidth]{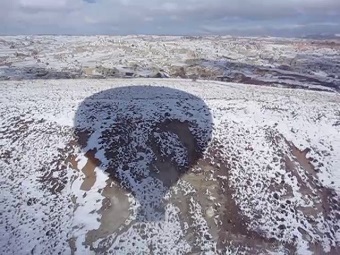}\vspace{1pt}
    \includegraphics[width=1\linewidth]{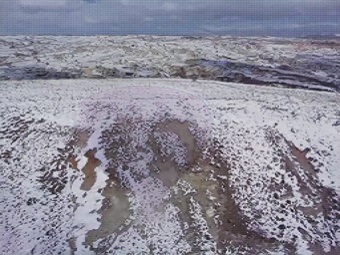}\vspace{1pt}
  \end{minipage}
  \vspace{-0.3cm}
  \caption{The visualization of shadow-removal results in a video. Note the frames are extracted every 100 milliseconds.}
\label{fig:video_removal}
\end{figure}

\textbf{Extension to video}
We apply our RIS-GAN to handle shadow videos by processing each frame in order. Figure \ref{fig:video_removal} presents the shadow-removal results for the frames every 100 milliseconds. From this Figure we can observe that the video shadow-removal results by applying image-level shadow removal approach to video directly are not good enough and there is still room for better improvement.

What's more, although our proposed RIS-GAN framework is designed for shadow removal, it is not limited to the shadow removal only. It is easy to be extended and applied to general image-level applications such as rain removal, image dehazing, intrinsic image decomposition, as well as other image style-transfer tasks.

\section{Conclusions}
In this paper, we have proposed a general and novel framework RIS-GAN to explore the residual and illumination for shadow removal. The correlation among residual, illumination and shadow has been well explored under a unified end-to-end framework. With the estimated negative residual image and inverse illumination map incorporating into the shadow refinement, we are able to get complementary input sources to generate a high-quality shadow-removal image. The extensive experiments have strongly confirmed the advantages of incorporating residual and illumination for shadow removal.

Our future work includes extending the current work to video-level shadow removal and applying the explored residual and illumination to solve the real challenging vision problems, such as image illumination enhancement.

\section{Acknowledgments}
This work was partly supported by Key Technological Innovation Projects of Hubei Province (2018AAA062), the National Natural Science Foundation of China (No. 61672390, No. 61972298, NO. 61902286, No.61972299, No.U1803262), the National Key Research and Development Program of China (2017YF-B1002600), China Postdoctoral Science Found (No. 070307), and the Joint laboratory Foundation of Xiaomi Company and Wuhan University. Chunxia Xiao is the corresponding author.


\bibliographystyle{aaai}
\bibliography{RIS-GAN} 

\end{document}